\documentclass[twocolumn,showpacs,preprintnumbers,amsmath,amssymb]{revtex4}
\usepackage{epsfig}
\usepackage{hyperref}

\newcommand{\ket}[1]{\ensuremath{| #1 \rangle}}
\begin{document}

\title{Bell inequalities and density matrix for polarization entangled photons out of a two-photon cascade in a single quantum dot}

\author{M. Larqu\'e}
\author{I. Robert-Philip}
\author{A. Beveratos}
\affiliation{CNRS - Laboratoire de Photonique et Nanostructures, Route de Nozay, F-91460 Marcoussis, FRANCE}
\date{\today}

\begin{abstract}
We theoretically investigate the joint photodetection probabilities of the biexciton-exciton cascade in single semiconductor quantum dots and analytically derive the density matrix and the Bell's inequalities of the entangled state. Our model includes different mechanisms that may spoil or even destroy entanglement such as dephasing, energy splitting of the relay excitonic states and incoherent population exchange between these relay levels. We explicitly relate the fidelity of entanglement to the dynamics of these processes and derive a threshold for violation of Bell's inequalities. Applied to standard InAs/GaAs self-assembled quantum dots, our model indicates that spontaneous emission enhancement of the excitonic states by cavity effects increases the fidelity of entanglement to a value allowing for violation of Bell's inequalities.
 
\end{abstract}
\pacs{
     42.50.Dv,  78.67.Hc, 81.07.Ta, 03.65.Ud 
     } 
 \maketitle

\section{Introduction}

Entangled photon pairs are an essential tool for quantum information science, ranging from quantum cryptography \cite{Ekert91}, to the realization of quantum relays \cite{Collins05} or quantum information processing \cite{Shor97, Grover97}. Quantum relays are probably one of the most advanced application using entanglement and have been implemented in real world quantum teleportation setups \cite{Ursin04, Landry07} or entanglement swapping demonstrations \cite{Pan98, Halder07}. In these experiments, entangled photons were obtained by parametric downconversion, but other sources based on 4-wave mixing are also investigated. Such non-linear sources of entanglement can combine narrow spectral bandwidths with a maximal generation rate \cite{Halder07, Kumar05, Fulconis07}. However, although these sources may be very useful and easy to implement, they always suffer from 
the Poissonian statsitics of the emitted photons pairs leading to multipair emission, which decreases the visibility of entanglement \cite{Scarani05}. The need to minimize the likelihood of producing multiple photon pairs forces these sources to be operated at low rates of photon pair generation per coherence length or excitation pulse (usually lower than 0.1). On the other hand, a deterministic source of entangled photons would make it possible to suppress these multipair events and to create light pulses with increased probability of containing a single photon pair, hence rendering all the above mentioned protocols much more efficient. From this point of view, sources based on the cascade emission from a single dipole (such as a single atom or a single quantum dot) may be a good candidated. In such system, the single dipole can be described as a four-level system emitting a single pair of photons upon each excitation cycle. For example, in self-assembled quantum dots, this cascade emission involves a biexciton, which consists of two-electron-hole pairs trapped in the dot with opposite angular momentum and which decays radiatively through two relay bright exciton \cite{Benson2000, Moreau2001}. This decay may release time-bin entangled photons \cite{Simon2005}, or polarization entangled photons \cite{Benson2000}. Time-bin entangled photons can also be obtained from two successive indistinguishable single photons \cite{Larque2008}. The origin of polarization entanglement here resides in the existence of two radiative decay paths with different polarizations which are otherwise indistinguishable. However, in such solid-state single emitters, polarization entanglement is spoiled by the anisotropic exchange interaction caused by in-plane anisotropy of the exciton wave function \cite{Gammon96, Bayer02}; such electron-hole exchange interaction lifts the excitonic states degeneracy and provides information about which pathways the two photons were released along via the energy of the emitted photons \cite{Santori2002}. Reducing the excitonic energy splitting within the radiative linewidth of the excitonic levels (by spectral filtering \cite{Akopian2006},  use of external magnetic \cite{Stevenson2006} or electric \cite{Gerardot2007} field, growth optimization \cite{Seguin2005}...) can in principle allow us to erase the which path information due to the excitonic fine structure and recover entanglement. However, dephasing interactions with the solid-state environment (for example through collisions with phonons and electrostatic interactions with fluctuating charges located in the dipole vicinity \cite{Berthelot2006}) may also degrade the strong correlations between the polarization of the two photons. Moreover, any incoherent mechanisms inducing a population exchange between the excitonic levels (such as transitions through the dark states or spin flip processes) may deteriorate the visibility of entanglement.

This paper theoretically investigates the joint photodetection probabilities in the biexciton cascade and analytically derive the density matrix as well as a non-optimal but nevertheless interesting entanglement witness based on the CHSH inequalities. Several incoherent process have been taken in account such as exciton energy splitting, incoherent population exchange between the excitonic levels and cross-dephasing between these two relay states. The following part of the paper begins by defining a Hamiltonian of a four-level system interacting with a solid-state environment and subject to incoherent population exchange between the two relay levels. We derive from such Hamiltonian a time evolution equation of the system excited on its upper state and derive the joint photodetection probability. In section \ref{Theory}, we quantify the entanglement of the photons produced by deriving an analytical expression of the Clauser-Horne-Shimony-Holt (CHSH) inequality as a function of the different dynamical parameters of the four-level system, as well as the density matrix corresponding to the biexciton cascade. We then stress in section \ref{Cavityeffects} the necessity to make use of the Purcell effect \cite{Gerard1998}, in order to violate Bell's inequalities from the cascade emission in self-assembled quantum dots.

\section{Theoretical framework}

\subsection{The four-level system}

\begin{figure}[!h]
\begin{center}
\includegraphics[scale=0.3]{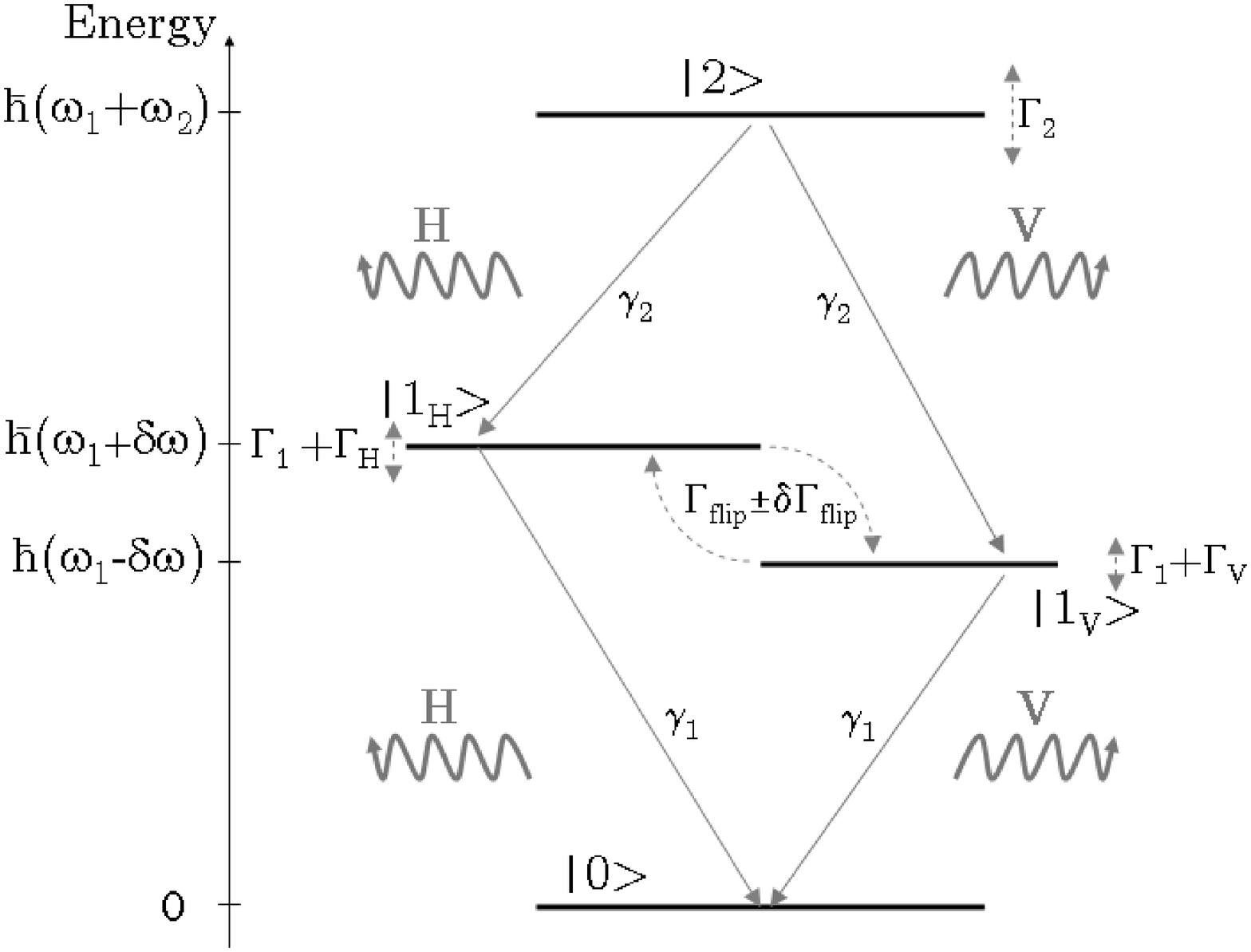}
\caption{Schematic description of the two-photon cascade in a typical four-level system with an energy splitting $2\hbar \delta \omega$
of the relay level, yielding two colinearly polarized photons (either $H$ or $V$).} \label{Fig1}
\end{center}
\end{figure}

In the cascade emission from a four-level system, the decay paths involve two radiative transitions, one from an upper level $|2\rangle$ to an intermediate state $|1_H\rangle$ or $|1_V\rangle$ and the other from these relay states to the ground state $|0\rangle$ (see Fig.\ref{Fig1}). The energies of these levels $|2\rangle$, $|1_H\rangle$ and $|1_V\rangle$ are respectively denoted $\hbar (\omega_1+ \omega_2)$, $\hbar (\omega_1 +\delta\omega)$ and $\hbar (\omega_1 -\delta\omega)$. We will futher assume that this $\{|2\rangle, |1_H\rangle,|1_V\rangle, |0\rangle\}$ basis corresponds to the eigen basis of the quantum dot, with therefore an excitonic energy splitting $2\delta \omega$ but no coherent coupling between the two excitonic eigenstates \cite{Santori2002}. Radiative transitions from the biexciton in such basis release colinearly polarized photons with linear polarization denoted $H$ and $V$ (see Fig. \ref{Fig1}). In the ideal case ($\delta\omega=0$), the four-level system relaxes, generating the maximally entangled two-photon state: 
\begin{equation}
|\Phi^+\rangle = \frac{1}{\sqrt{2}}(|H, \omega_1\rangle|H, \omega_2\rangle + |V, \omega_1\rangle|V, \omega_2\rangle)
\label{eq:phiplus}
\end{equation}

\noindent by cascade emission \cite{Aspect82, Benson2000}. The phase difference between the two component states $|H, \omega_1\rangle|H, \omega_2\rangle$ and $|V, \omega_1\rangle|V, \omega_2\rangle$ is null, as determined by the angular momenta of the different involved levels and the Clebsch-Gordan coefficients \cite{Fry73}. Unfortunately, in realistic two-level systems (such as single quantum dots for example), the relay levels are split ($\delta\omega \neq 0$). Furthermore, relaxation mecanisms between the two relay states $|1_H\rangle$ and $|1_V\rangle$ can occur (for example from spin flip processes). They will be accounted for by two phenomenological decay rates $\Gamma_{flip} \pm \delta\Gamma_{flip}$, that will be latter supposed to be equals (which is a good approximation for a small excitonic energy splitting). 

In addition, the relay levels and the upper level may be subject to sudden, brief and random fluctuations of their energies without population exchange (arising, for example, from collisions with thermal phonons). In our model, the ground level $|0\rangle$ is chosen as the reference in energy and phase. Dephasing of the upper level $|2\rangle$ is described by the dephasing rate $\Gamma_2$. On the two relay levels $|1_H\rangle$ and $|1_V\rangle$, we distinguish two dephasing processes without population exchange between these relay levels: (1) dephasing processes that occur simultaneously and attach the same information on the phase and energy of these two levels with a dephasing rate denoted $\Gamma_1$ and (2) dephasing processes that do not affect identically the two relay levels and whose impact depends on the polarization of the excitonic states. These last processes will be described by polarization-dependent dephasing rates $\Gamma_H$ and $\Gamma_V$. The cross-dephasing between the two relay states is therefore $\Gamma = \Gamma_H + \Gamma_V$. This model includes all possible dephasing processes without population modifications that may occur.

\subsection{Dynamics of the four level system}

In order to account for the open nature of the four-level system (resulting from its coupling with the phonon and the photon reservoirs for example), we describe the time evolution of the density operator $\rho$ by means of the following master equation in the Lindblad form \cite{Lindblad1975}:

\begin{equation} 
\frac{d\rho}{dt} = -[iH, \rho] + (\mathcal{L}_r + \mathcal{L}_d + \mathcal{L}_{flip})\rho 
\label{eq:MasterEquation}
\end{equation}

\noindent In the previously described eigen basis $\{|2\rangle,|1_H\rangle,|1_V\rangle,|0\rangle\}$ of the four-level system, the hamiltonian $H$ has the form:

\begin{equation}
H = (\omega_1-\delta\omega)|1_V\rangle \langle 1_V|
	+(\omega_1+\delta\omega)|1_H\rangle \langle 1_H|
	+(\omega_1+\omega_2)|2\rangle \langle 2|
\end{equation}

\noindent The Lindblad operators include three contributions. The first one describes the interaction of the emitter with the electromagnetic field by emitting photons, whenever it undergoes a transition from its upper state to the relay levels or from the relay levels to the ground state.  This radiative relaxation is accounted for by the following Liouvillian:

\begin{equation}
\mathcal{L}_r = \sum_{p=H,V} (\frac{\gamma_1}{2}
\mathcal{L}(|0\rangle\langle 1_p|)+\frac{\gamma_2}{2}
\mathcal{L}(|1_p\rangle\langle 2|))
\end{equation}

\noindent where $\gamma_1$ and $\gamma_2$ are respectively the radiative decay rates between the relay states and the ground state and between the upper level and the relay levels. We assume that these decay rates do not depend on the decay path the photons were released along. $\mathcal{L}(D)\rho = 2D\rho D^\dagger - D^\dagger D \rho - \rho D^\dagger D
$ is the Lindblad operator. The second contribution $\mathcal{L}_d$ is related to dephasing processes and reads:

\begin{eqnarray}
\mathcal{L}_d &=& \Gamma_2 \mathcal{L}(|2\rangle\langle 2|) 
	+ \sum_{p=H,V} \Gamma_p \mathcal{L}(|1_p\rangle\langle 1_p|) \nonumber\\
	&& + \Gamma_1 \mathcal{L}(|1_H\rangle\langle 1_H|+|1_V\rangle\langle 1_V|)
\end{eqnarray}

\noindent This Liouvillian includes phenomenologically any dephasing effect (e.g. phonons) occuring on the levels of the dot without population transfers as described previously. The last contribution $\mathcal{L}_{flip}$ accounts for the incoherent coupling between the two relay states: 

\begin{eqnarray}
\mathcal{L}_{flip} &=& \alpha_P\mathcal{L}(|1_H\rangle\langle 1_V|+|1_V\rangle\langle 1_H|) \nonumber\\
&& +\beta_P\mathcal{L}(i(|1_H\rangle\langle 1_V|+|1_V\rangle\langle 1_H|)) \nonumber\\
&& +\beta_Q\mathcal{L}(|1_V\rangle\langle 1_H|-|1_H\rangle\langle 1_V|) \\ 
&& +\alpha_Q\mathcal{L}(i(|1_V\rangle\langle 1_H|-|1_H\rangle\langle 1_V|)) \nonumber
\end{eqnarray}

\noindent The phenomenological rate $\Gamma_{flip}$ between the two relay states $|1_H\rangle$ and $|1_V\rangle$ appears to be twice the sum of the different rates $\alpha_i$ and $\beta_i$ ($i=P, Q$) involved in this equation. The rate $\delta \Gamma_{flip}$ expresses likewise as : $2(\alpha_Q - \alpha_P + \beta_Q - \beta_P)$. These rates simulate any unspecified process inducing an incoherent interaction between the two relay levels with population exchange before radiative relaxation. $\delta \Gamma_{flip}$ accounts for assymetry of these processes. These non-radiative processes may include spin-flip processes and transitions through dark states (assuming that the probability for the source to be in these dark-states is small compared to the probabilities related to the optically active states). \\

In accordance with all these assumptions, the time evolution of the four-level system can be decomposed by use of master equation (\ref{eq:MasterEquation}) in a set of differential equations, which reduces for the purpose of this paper to:
\begin{equation}
\frac{dV}{dt}=A V
\label{eq:evolution}
\end{equation}
\noindent In this equation, $V$ is a vector composed of the following mean values:
\begin{equation}
V = \left( \begin{array}{c} Tr[S_\Delta\rho] \\ Tr[S_Q\rho] \\ Tr[S_P\rho] \end{array} \right) \\
\label{eq:defV}
\end{equation}

\noindent The operator $S_\Delta=|1_H\rangle \langle 1_H|-|1_V\rangle \langle 1_V|$ is related to population difference between the two excitonic relay states. The two other Pauli's matrices $S_Q=\i|1_V\rangle\langle 1_H|-\i|1_H\rangle\langle 1_V|$ and $S_P=|1_H\rangle\langle 1_V|+|1_V\rangle\langle 1_H|$ correspond to the quadratures of the dipole between these two relay states. The $A$ matrix is given by:
\begin{eqnarray}
A &=& \left( \begin{array}{c c c}
	- \gamma_1 - 2\Gamma_{flip} & 0 & 0\\ 
	0 & - \mu_Q & 2\delta\omega  \\ 
	0 & -2\delta\omega & -\mu_P 
	\end{array}\right)
\label{eq:setofsix} 
\end{eqnarray}

\noindent The decay constants $\mu_{P/Q}$ are equal to $\gamma_1 + \Gamma_{flip} \pm \delta\Gamma_{flip} + \Gamma$.  \\

For further reference, we define the matrix transformation $M(U)$ of $V$, where $U$ is an arbitrary unitary transformation of the excitonic levels of the source (letting the upper and fundamental states unchanged), by
\begin{equation}
M(U)V(t)=\left(\begin{array}{c} 
	Tr[U S_\Delta U^\dagger \rho(t)] \\
	Tr[U S_Q U^\dagger \rho(t)] \\
	Tr[U S_P U^\dagger \rho(t)]
	\end{array}\right)
\label{eq:defM}
\end{equation}
\noindent M(U)V(t) are the mean values (\ref{eq:defV}) measured at time $t$ under the transformed basis $\{|2\rangle,U|1_H\rangle,U|1_V\rangle,|0\rangle\}$.

\subsection{Joint photodetection probability}

\begin{figure}[!h] 
\begin{center}
\includegraphics[scale=0.3]{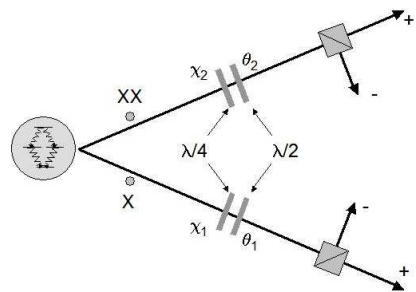}
\caption{Experimental setup for measuring CHSH or reconstructing the density matrix} 
\label{Fig2}
\end{center} 
\end{figure}

Violation of Bell inequalities as well as the reconstruction of the density matrix is experimentally obtained by measuring the joint photon detection probabilities P$_{\pm,\pm}$ on the output of a binary polarization analyzer such as a polarizing beamsplitter. Since several points of the Bloch sphere have to be measured\cite{Aspect82}, a quater-wave plate followed by a half-wave are inserted in the photons path (see fig.\ref{Fig2}). The exciton and biexciton photon are spectrally separated by means of optical filters and send through the optical path denoted $i$ ($i$=1 or 2 for the exciton and biexciton respectively)

The fast axis of the quarter-(resp. half-)wave plate is rotated by an angle $\chi_i$ (resp. $\theta_i$) with respect to the horizontal polarization direction defined by the optical table. By applying the projection theorem, measuring $+1$ in the optical setup $i$ corresponds to the detection of a photon $i$ emitted by the source with the polarization $\Lambda(\theta_i,\chi_i)^\dagger|H\rangle$ where $\Lambda(\theta,\chi)$ describes the transformation of the polarization basis $\{H,V\}$ when a photon successively propagates through a quarter- and a half-wave plates rotated by the angles $\theta$ and $\chi$.
\begin{equation}
\Lambda(\theta,\chi)=R(\theta)T(\pi)R(\chi-\theta)T(\pi/2)R(-\chi)
\end{equation}
\noindent where $R(x)$ is the rotation matrix and $T(r)$ is the Jones matrix of a retarder plate.
\begin{equation}
R(x) = \left( \begin{array}{c c} cos(x) & sin(x) \\ -sin(x) & cos(x) \end{array}\right), \
T(r) = \left( \begin{array}{c c} 1 & 0 \\ 0 & e^{-\i r} \end{array}\right)
\end{equation}
\noindent In the following, for the sake of clarity we will denote $\Lambda(\theta,\chi)^\dagger|1_H\rangle$ the superposition of the source's states $|1_H\rangle$ and $|1_V\rangle$ which analytically corresponds to the same transformation $\Lambda(\theta,\chi)^\dagger|H\rangle$ of the photonic state $|H\rangle$.

Experimentally one measures the joint photodetection probabilities $P_{\pm, \pm}^{det}(\theta_2,\chi_2,\theta_1,\chi_1)$ of the first photon and second photon in channels $\pm$ of their respective optical setups with each retarder plate rotated by $\theta_i$ and $\chi_i$.

The source is pumped at time $t=0$ from its ground state to the excited state $|2\rangle$ with a laser pulse shorter than the lifetime $1/(2\gamma_2)$ of the upper state. We will futher postselect joint photodetection events corresponding to a sequential detection of the biexcitonic photon and then of the excitonic photon during one excitation cycle. In this context, the probability of joint photodetection $P_{+,+}^{det}(\theta_2,\chi_2,\theta_1,\chi_1)$ is proportional to the emission probability $P_{+,+}(\theta_2,\chi_2,\theta_1,\chi_1)$ of a pair of photons with respective polarization orientation $\Lambda(\theta_i,\chi_i)^\dagger|H\rangle$, at respective energies $\hbar \omega_2$ and $\hbar \omega_1$, assuming that the source is in state $|2\rangle$ at time $t=0$. This radiative transition probability can be regarded as the product of two probabilities: the probability of emission of the first photon with polarization $\Lambda(\theta_2,\chi_2)^\dagger|H\rangle$, multiplied by the conditional probability of radiative transition from the relay levels to the ground state with emission of a photon polarized along $\Lambda(\theta_1,\chi_1)^\dagger|H\rangle$. This amounts in considering the photon cascade as a two-step process and applying the quantum-measurement projection postulate. First the photon at energy $\hbar \omega_2$ and polarized along $\Lambda(\theta_2,\chi_2)^\dagger|H\rangle$ is detected at time $t_2$, which projects the emitter on the superposition $\Lambda(\theta_2,\chi_2)^{*\dagger}|1_H\rangle$ of the exciton states $|1_H\rangle$ and $|1_V\rangle$. Secondly, the superposition state evolves in time until the detection of the second photon at energy $\hbar \omega_1$ at time $t_1+t_2$. Consequently, this conditional probability will be related to the population in the superposition $\Lambda(\theta_1,\chi_1)^\dagger|1_H\rangle$ at time $t_2+t_1$, knowing that the intermediate levels were in the superposition $\Lambda(\theta_2,\chi_2)^{*\dagger}|1_H\rangle$ at time $t_2$. All these probabilities are integrated over the photodetection time window. 

The population at time $t_1+t_2$ in the superposition $|1_H(\theta_1)\rangle$ can be expressed as $[e^{-\gamma_1 t_1}+ \langle S_\Delta\rangle(t_2+t_1 | t_2)]/{2}$ where $\langle S_\Delta\rangle(t_2+t_1 | t_2)$ is the first value of the vector $V$ of Eq. \ref{eq:defV} measured under the transformation of Eq. \ref{eq:defM} with $U=\Lambda(\theta_1,\chi_1)^\dagger$, after a free evolution during the time $t_1$ (Eq. \ref{eq:evolution}) with the assumption of the inital state $V^{init}$ corresponding to the excitonic state $\Lambda(\theta_2,\chi_2)^{*\dagger}|H\rangle$ at time $t_2$. Thus by defining the vector $V_0=\{1,0,0\}$ which corresponds to the values of $V$ measured in the eigenbasis with the source in the state $|1_H\rangle$, it follows:
\begin{eqnarray}
V^{init} &=& M(\Lambda(\theta_2,\chi_2)^{*\dagger})^{-1} V_0 \nonumber\\
V^{measured} &=& M(\Lambda(\theta_1,\chi_1)^\dagger) e^{At} V^{init} \nonumber\\
\langle S_\Delta\rangle(t_1 | 0) &=& \langle S_\Delta\rangle(t_2+t_1 | t_2) \\
	&=& [M(\Lambda(\theta_1,\chi_1)^\dagger) e^{At} M(\Lambda(\theta_2,\chi_2)^{*\dagger})^{-1}]_{11} \nonumber
\end{eqnarray}

\noindent where $[\ldots]_{ij}$ denotes the matrix element on row $i$ and column $j$. The probability $P_{+, +}(\theta_2,\chi_2,\theta_1,\chi_1)$ can therefore be written as follows:
\begin{eqnarray}
P_{+,+}(\theta_2,\chi_2,\theta_1,\chi_1) = \int_0^{+\infty} \gamma_2 e^{-2 \gamma_2 t_2} dt_2 \nonumber\\
	\times \int_0^{+\infty} \frac{\gamma_1}{2} (e^{-\gamma_1 t_1}+\langle S_\Delta\rangle(t_1|0) dt_1
\label{eq:probaplusplus_int}
\end{eqnarray}

Upon integration, this probability reads in the particular case $\chi_1=\chi_2=0$:

\begin{widetext}
\begin{equation}
P_{+,+}(\theta_2, \theta_1) = \frac{1}{4}[1+
	\frac{\gamma_1}{\gamma_1+2\Gamma_{flip}} cos(4\theta_1)cos(4\theta_2)
	+ \frac{\gamma_1 (\gamma_1 + \Gamma_{flip} + \Gamma - \delta\Gamma_{flip})}{(2\delta\omega)^2 + (\gamma_1 + \Gamma_{flip} + \Gamma)^2 - (\delta\Gamma_{flip})^2} sin(4\theta_1)sin(4\theta_2)] 
\label{eq:probaplusplus}
\end{equation}
\end{widetext}

For a perfect quantum dot, $P_{+,+}(0,0)$ tends toward $1/2$ as expected.

\section{Quantifying two-photon entanglement and density matrix}\label{Theory}

Entanglement can be quantified by several means like measurement of the concurrence, tangle of the density matrix or entanglement witness operators. A non optimal entanglement witness, but nevertheless experimentally simple to measure is the Bell inequality under the CHSH form which discriminate between states that can be explained by a Local Hidden Variable Model (LHVM) or not. The possible violation of Bell inequalities is experimentally easy to verify by measuring the fringes visibility \cite{Marcikic2004} of two-photon coincidences as a function of ($\theta_1-\theta_2$) whereas other measurements need the experimental knowledge of the density matrix. Hence we shall first derive the analytical form of the CHSH inequality, then generalize the result to the derivation of the density matrix and one possible entanglement witness \cite{Hyllus2005} by use of the Peres criterion \cite{Peres1996}.

\subsection{Violation of Bell's inequalities}

The CSHS inequality is calculated by measuring the correlation coefficient for four sets of properly chosen angles of a half-wave plate, and therefor the angles $\chi_i$ referring to the quarter wave plate are set to zero and omitted in the rest of this subsection. From the expression of Eq. \ref{eq:probaplusplus} one deduces all the probabilities $P_{\pm, \pm}(\theta_2,\theta_1)$ and compute analytically in a straightforward manner the correlation coefficient of the form:
\begin{eqnarray} 
E(\vec{\theta_2}, \vec{\theta_1}) &=& P_{+, +}(\vec{\theta_2}, \vec{\theta_1}) 
	+ P_{-, -}(\vec{\theta_2}, \vec{\theta_1}) \nonumber\\
	&& - P_{-,+}(\vec{\theta_2}, \vec{\theta_1}) 
	- P_{+, -}(\vec{\theta_2}, \vec{\theta_1}) 
\label{eq:CorrelCoeff1}
\end{eqnarray}

The generalized Bell's inequality in the Clauser-Horne-Shimony-Holt (CHSH) formulation \cite{CHSH} is expressed as a combination of such correlations functions as:
\begin{eqnarray}
S(\theta_2, \theta_2', \theta_1, \theta_1') &=& E(\vec{\theta_2}, \vec{\theta_1}) - E(\vec{\theta_2'}, \vec{\theta_1}) \nonumber\\
	&& + E(\vec{\theta_2}, \vec{\theta_1'}) + E(\vec{\theta_2'}, \vec{\theta_1'})
\label{eq:CorrelCoeff2}
\end{eqnarray}

\noindent which, for classically correlated states, is bounded by $|S| \leq 2$. In the case of an ideal entangled source, the maximum value of the CHSH coefficient $S$ is obtained for every set of polarization directions of each analyzer verifying $\theta_2 =x+\pi/16$ and $\theta_2'=x+3\pi/16$; $\theta_1 =x$ and $\theta_1'=x+\pi/8$, where $x$ is an arbitrary rotation of both half-wave plates. In this context and under the assumption of $\delta\Gamma_{flip}=0$, the CHSH parameter $S$ is given by the formula:
\begin{equation}
S=\sqrt{2}\left( \frac{\gamma_1}{\gamma_1 + 2\Gamma_{flip}} + \frac{\gamma_1(\gamma_1 + \Gamma_{flip} + \Gamma)}{\left(\gamma_1 + \Gamma_{flip} + \Gamma \right)^2 + (2\delta\omega)^2} \right)
\label{CHSH_S}
\end{equation}
\noindent which is, as expected, independent of the arbitrary rotation $x$. Violation of Bell's inequalities implies $S>2$.

\subsection{Density Matrix}

In the above section, the Bell inequalities have been derived form the coincidence probabilities when the exciton and biexciton are detected in a linear basis. We will now exploit the more general expression of the joint photodetection probability $P_{+,+}(\theta_2,\chi_2,\theta_1,\chi_1)$ obtained upon integration of Eq. \ref{eq:probaplusplus_int}. 
By definition of the density matrix, this probability can also be expressed as 
\begin{widetext}
\begin{equation}
P_{+,+}^{\rho}(\theta_2,\chi_2,\theta_1,\chi_1) = \langle H_{XX}H_{X}| (\Lambda(\theta_2,\chi_2) \otimes \Lambda(\theta_1,\chi_1)) \cdot \underline\rho \cdot (\Lambda(\theta_2,\chi_2) \otimes \Lambda(\theta_1,\chi_1))^\dagger |H_{XX}H_{X}\rangle \nonumber
\end{equation}
\end{widetext}
\noindent where $\underline\rho$ is the density matrix of the pair of photon in the basis $\{H_{XX}H_{X},H_{XX}V_{X},V_{XX}H_{X},V_{XX}V_{X}\}$. By identifying $P_{+,+}^{\rho}(\theta_2,\chi_2,\theta_1,\chi_1) = P_{+,+}(\theta_2,\chi_2,\theta_1,\chi_1)$ for $16$ well chosen set of four angles $(\theta_2,\chi_2,\theta_1,\chi_1)$ we construct a linear system of $16$ independent equation whose unknown variables are the $16$ real values of $\underline\rho$. In this way, we simply reconstruct the density matrix from the joint photodetection probabilities\cite{White2001} and obtain a theoretical value of $\underline\rho$. Same holds for an experimental approach.
%
The calculated density matrix is hence :
\begin{equation}
\underline\rho = \left( \begin{array}{cccc}
	\alpha & 0 & 0 & d - \i c_1 \\
	0 & \beta & c_2 & 0 \\
	0 & c_2 & \beta & 0 \\
	d + \i c_1 & 0 & 0 & \alpha
	\end{array}\right)
\end{equation}
\noindent where
\begin{eqnarray}
\alpha &=& \frac{1}{2} \frac{\gamma_1+\Gamma_{flip}}{\gamma_1+2\Gamma_{flip}} \nonumber\\
\beta &=& \frac{1}{2} \frac{\Gamma_{flip}}{\gamma_1+2\Gamma_{flip}} \nonumber\\
d &=& \frac{1}{2} \frac{\gamma_1(\gamma_1+2\Gamma+\Gamma_{flip})}{(2\delta\omega)^2 + (\gamma_1 + \Gamma_{flip} + \Gamma)^2 - (\delta\Gamma_{flip})^2}\\
c_1 &=& \frac{1}{2} \frac{\gamma_1 \delta\omega}{(2\delta\omega)^2 + (\gamma_1 + \Gamma_{flip} + \Gamma)^2 - (\delta\Gamma_{flip})^2}\nonumber\\
c_2 &=& \frac{1}{2} \frac{\gamma_1 \delta\Gamma_{flip}}{(2\delta\omega)^2 + (\gamma_1 + \Gamma_{flip} + \Gamma)^2 - (\delta\Gamma_{flip})^2}\nonumber
\end{eqnarray}

Note that for a perfect quantum dot ($\delta \omega \rightarrow 0$ $\Gamma_{flip}\rightarrow 0$, $\Gamma \rightarrow 0$), $\rho$ tends, as expected, toward the $|\Phi^+\rangle$ Bell state: $\underline\rho \rightarrow |\phi^+ \rangle\langle \phi^+|$.

\subsection{Quantum dot spectroscopy from quantum tomography}

An interesting feature of the analytical form of the density matrix arises from the fact that once computed experimentally, one can deduce all the quantum dot parameters provided $\gamma_1$ is measured independently. They are expressed as a function of the density matrix elements :

\begin{eqnarray}
\delta \omega & = & \gamma_1 \frac{c_1}{4(d^2+c_1^2-c_2^2)} \nonumber\\
\delta \Gamma_{flip} & = & \gamma_1 \frac{c_2}{2(d^2+c_1^2-c_2^2)} \\
\Gamma_{flip} & = & \gamma_1 \frac{2\beta}{4\beta-1} \nonumber \\
\Gamma & = & \gamma_1 \frac{d(1-2d-4\beta+4d\beta) + 2(c_1^2-c_2^2)(2\beta-1)} {4(d^2+c_1^2-c_2^2)(4\beta-1)} \nonumber
\end{eqnarray}

\subsection{Entanglement witness}

Apart of the CHSH inequality other entanglement witnesses can be constructed and following \cite{Hyllus2005} we define an entanglement witness as $Tr[W\underline\rho]$ where $W$ is an operator. In the case where $W$ is an optimal witness, the above mentioned quantity is negative if $\rho$ is an entangled state. As proposed in \cite{Peres1996} we define the partial transpose $\rho_0^{T_2}$ of an arbitrary density matrix $\rho_0$ as follows :
\begin{eqnarray}
\rho_0 &=& \sum_{i,j,k,l=H,V} \underline\rho_{0\ i,j,k,l} |ij \rangle\langle kl| \nonumber\\
\rho_0^{T_2} &=& \sum_{i,j,k,l=H,V} \underline\rho_{0\ k,j,i,l} |ij \rangle\langle kl|
\end{eqnarray}

As demonstrated in \cite{Peres1996}, if $\underline\rho^{T_2}$ has a negative eigenvalue $\lambda$ associated to the eigenvector $|\nu\rangle$ then the density matrix $\underline\rho$ represents an entangled state. Thus defining $W=|\nu\rangle\langle\nu|^{T_2}$ we have
\begin{equation}
Tr[W\underline\rho_0] = Tr[|\nu\rangle\langle\nu| \underline\rho_0^{T_2}] = \lambda < 0
\end{equation}

In our case we choose for $\rho_0$ the density matrix $|\phi^+ \rangle\langle \phi^+|$ toward which the biexciton density matrix $\underline\rho$ of our model tends to. This gives a non optimal witness but already less demanding than the Bell's inequalities and with a simple analytical form:
\begin{eqnarray}
W &=&  \left( \begin{array}{cccc}
	0 & 0 & 0 & -1/2 \\
	0 & 1/2 & 0 & 0 \\
	0 & 0 & 1/2 & 0 \\
	-1/2 & 0 & 0 & 0
	\end{array}\right) \\
Tr[W\underline\rho] &=& \beta - d \nonumber\\
	&=& \frac{1}{2} \frac{\Gamma_{flip}}{\gamma_1+2\Gamma_{flip}} \\
	&&  - \frac{1}{2} \frac{\gamma_1(\gamma_1+2\Gamma+\Gamma_{flip})}{(2\delta\omega)^2 + (\gamma_1 + \Gamma_{flip} + \Gamma)^2 - (\delta\Gamma_{flip})^2} \nonumber
\end{eqnarray}

\section{\label{Cavityeffects} Restoration of entanglement through cavity effects}

In the above section we have analytically derived the CHSH inequality as well as the density matrix for the biexciton-exciton cascade emission from a single semiconductor quantum dot. Although we could discuss on the density matrix as a function of the internal parameters of the QD, we choose to discus the CHSH inequality since it is an intuitive entanglement witness with a simple experimental realization.
Equation \ref{CHSH_S} indicates that polarization entanglement in the cascade emission from a biexciton in a self-assembled quantum dot may be affected by the relative contribution of three processes with respect to the exciton radiative lifetime $1/\gamma_1$:  the mutual coherence between the two non-degenerate excitonic levels described by a cross-dephasing time $1/\Gamma$, the excitonic energy splitting giving rise to quantum beats with a time period $2\pi/2\delta \omega$ and the incoherent population exchange between the two bright excitons with a decay time $1/\Gamma_{flip}$. Entanglement does not depend on the biexciton radiative rate ($\gamma_2$) and among all the dephasing processes taken into account in our model, only the cross-dephasing between the excitonic levels affects the visibility of entanglement. The analytical expression of $S$ given by (\ref{CHSH_S}) also confirms that polarization entanglement from the biexciton cascade in self-assembled quantum dots is exclusively affected by the dynamics and mutual coherence of the excitonic states.

For quantum dots with no excitonic energy splitting ($\delta \omega=0$) and in absence of cross-dephasing ($\Gamma = 0)$ and incoherent population exchange ($\Gamma_{flip}$=0), the $S$ quantity reaches its maximum value of $2\sqrt{2}$ and the photons emerging from the biexcitonic cascade are maximally entangled \cite{Benson2000}. Conversely, for quantum dots whose excitonic states are splitted and which are affected by spin-dependent dephasing mechanisms and incoherent population exchange between the exciton bright states, the $S$ parameter rapidly decreases so that the two photons emitted are only partially entangled or even only correlated in one preferred polarization basis corresponding to the polarization eigenbasis of the dot \cite{Santori2002}.

\begin{figure}[!h] 
\begin{center}
\includegraphics[scale=0.3]{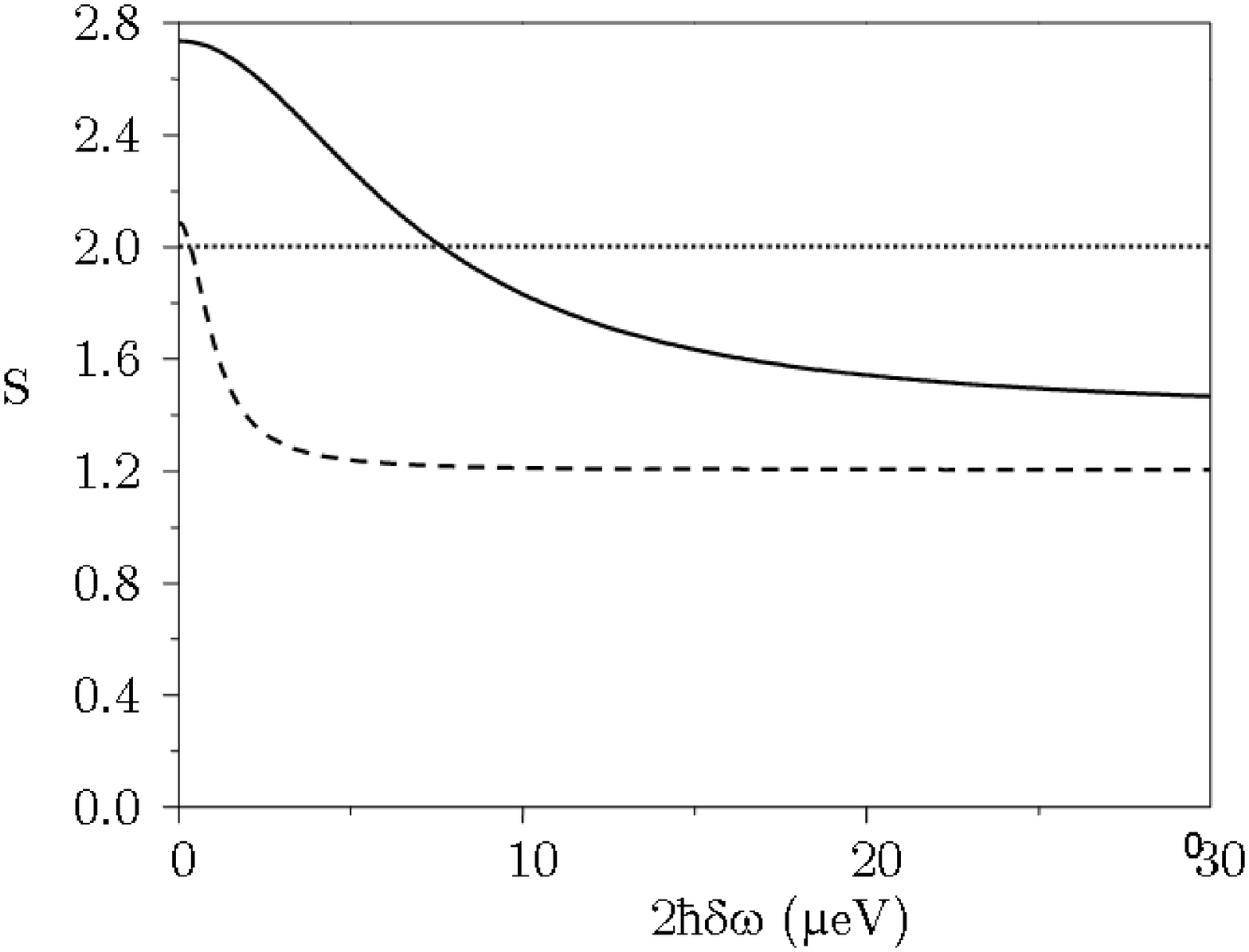}
\caption{CHSH inequality as a function of the energy splitting of the exciton line, for a single quentum dot in bulk material (dashed line) and subject to a Purcell effect with $F=10$ (continuous line). Dotted line corresponds to the classical limit of 2. For these two curves, $T_1=1/\gamma_1=1$ ns, $1/\Gamma_{flip}=10$ ns and $1/\Gamma=2$ ns.} 
\label{Fig3}
\end{center} 
\end{figure}

As an example, the characteristic excitonic lifetimes $1/\gamma_1$ of InAs quantum dots embedded in GaAs are typically on the order of 1 ns \cite{GerardAPL96} and the excitonic energy splitting $2\hbar \delta\omega$ is of the order of few $\mu$eV \cite{Langbein2004} corresponding to quantum beat periods lower than few hundreds ps. Numerous observations also indicate that the exciton spin relaxation is quite negligible on the timescale of the exciton lifetime and may reach values of about 10 ns or even higher \cite{Paillard2001, Kowalik2008}. The mutual coherence time $1/\Gamma$ is likely to be longer than few ns \cite{Hudson2007}, since it shall involve hypothetical spin-dependent dephasing processes. These typical values indicate that the main ingredient affecting entanglement is the excitonic energy splitting; they imply that in an experimental setup involving bare InAs quantum dots, the $S$ quantity is lower than 2 and tests of the Bell's inequalities on the two photons emerging from the biexciton cascade will not lead to any violation of the CHSH inequality (see dashed line on Fig. \ref{Fig3}). Even for relatively small exciton energy splitting ($2\hbar \delta\omega$ higher than few $\mu eV$), the $S$ value tends to 1.2, a value significantly lower than the $S=\sqrt{2}$ limit of perfectly correlated photons without any hidden variables. The incoherent population exchange between the excitonic relay levels destroys entanglement and the emitted photons are a statistical mixture of $\left\{\ket{HH},\ket{VV},\ket{HV},\ket{VH}\right\}$ states. Even for bare quantum dots with no exciton energy splitting, entanglement is spoiled by cross-dephasing and incoherent population exchange between the two bright excitonic states: the maximum value of $S$ on Fig. \ref{Fig3} for such quantum dots reaches only 2.06, a value very close to the classical limit of 2. 

\begin{figure}[!h] 
\begin{center}
\includegraphics[scale=0.3]{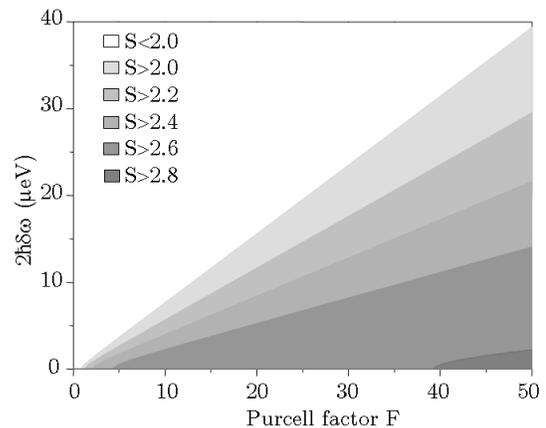}
\caption{CHSH inequality as a function of the energy splitting of the exciton line and its spontaneous emission exaltation $F$, for a single quantum dot with $T_1=1/\gamma_1=1$ ns in bulk material, $1/\Gamma_{flip}=10$ ns and $1/\Gamma=2$ ns.} 
\label{Fig4}
\end{center} 
\end{figure}

Nevertheless, restoration of entanglement and improvement of its visibility can be achieved by reducing the excitonic radiative lifetime of the quantum dot by a factor of $F$ through its introduction in a resonant microcavity and the exploitation of the Purcell effect \cite{Troiani}: by making the excitonic spontaneous emission faster than the quantum beats period ($F\gamma_1 \gg 2\delta \omega$), the cross-dephasing time ($F\gamma_1 \gg \Gamma$) and the decay time of incoherent excitonic population exchange ($F\gamma_1 \gg \Gamma_{flip}$), it should be possible to preserve the quantum correlations between the two recombination paths. We consider here that both excitonic transitions releasing either $H$ or $V$-polarized photons are accelerated by cavity effects with the same spontaneous emission enhancement factor $F$. For dots subject to a spontaneous emission enhancement of its excitonic transition by a factor $F=10$ (see solid line on Fig. \ref{Fig3}), $S$ values higher than 2.6 should be achievable for null exciton energy splitting ($2\hbar\delta\omega=0$). In such microcavity source however, violation of Bell's inequalities ($S>2$) requires the use of quantum dots with an excitonic energy splitting smaller than 7 $\mu eV$. With a Purcell effect of $F=30$, $S$ reaches the value of 2.76 close to its maximum value of $2\sqrt{2}$ for dots with no exciton energy splitting, and Bell's inequalities are violated for quantum dots displaying an energy splitting up to 20 $\mu eV$ (see Fig. \ref{Fig4}).

Figure \ref{Fig4} shows the value of $S$ as a function of $2\hbar\delta\omega$ and $F$ for values of $\gamma_1$, $\Gamma$ and $\Gamma_{flip}$ considered above as typical of currently available quantum dots.  The results confirm that the main ingredient degrading entanglement is the exciton fine structure. However, reducing the exciton energy splitting within the exciton linewidth is not experimentally sufficient and hardly allows for violation of Bell's inequalities. Violation of the CHSH inequalities requires a combination of cavity effects enhancing the excitons spontaneous emission and techniques leading to a reduction of the exciton energy splitting (such as growth optimization \cite{Seguin2005} or use of external magnetic \cite{Stevenson2006} or electric \cite{Gerardot2007} field). For typically available quantum dots, a Purcell factor of the order of 10 exalting equally both excitons transitions, would be sufficient for reaching values of $S$ higher than the classical limit of 2. Yet, the generation of maximally-entangled photons ($S=2\sqrt{2}$) with a single quantum dot is precluded by all decoherence mechanisms such as cross-dephasing between the exciton states and incoherent population exchange between the two bright excitons. Maximally entangled states could however still be obtained out of non-maximally entangled states by use of entanglement purification \cite{Pan_Purification}.\\

\section{Summary and conclusion}

We have shown analytically that in the two-photon cascade from the biexciton in a single semiconductor quantum dot, solely the dynamics and coherence of the excitonic dipole governs the visibility of polarization entanglement. We have derived Bell inequalities under the CHSH form, as well as the density matrix of such a state. In bare quantum dots, polarization entanglement is spoiled not only by the energy splitting of the relay level but also by the incoherent population exchange and cross-dephasing between the two bright relay states. The use of a microcavity can restore the generation of polarization-entangled photons from the quantum dot: The presence of the microcavity enhances the spontaneous emission rate of the excitonic transition, so that emission of the second photon arises before any quantum beat, cross-dephasing or incoherent population transfer between the excitonic radiative states. For experimentally accessible regime, violation of Bell's inequalities can be achieved with real quantum dots, provided a small excitonic energy splitting (lower than few $\mu eV$) and a Purcell factor of the order of 10. Such Purcell factors and excitonic energy splitting have already been achieved, indicating that the possibility of realizing polarization-entangled photons with semiconductor quantum dots embedded in microcavities is totally accessible with available technology. \\

\noindent {\bf Acknowledgements:} Numerous helpful discussions
with I. Abram, S. Laurent, O. Krebs, P. Voisin, V. Scarani and F. Grosshans are gratefully
acknowledged. This work was partly supported by the NanoSci-ERA
European Consortium under project ``NanoEPR''. We also acknowledge
support of the ``SANDiE'' Network of Excellence of the European
Commission.


\end{document}